\documentclass[prb,twocolumn,superscriptaddress,showpacs]{revtex4}

\usepackage{amsmath}
\usepackage{amssymb}
\usepackage{amsfonts}

\usepackage{graphicx}
\usepackage{times}
\usepackage{pstricks}

\newcommand{\be}{\begin{equation} }
\newcommand{\ee}{\end{equation} }
\newcommand{\ba}{\begin{eqnarray} }
\newcommand{\ea}{\end{eqnarray} }

\newcommand{\NOT}{{\rm{NOT}}^{-,+}}
\newcommand{\pow}{d}

\usepackage{amsthm}

\begin{document}


\title{Metaplectic Anyons, Majorana Zero Modes, and their Computational Power}

\author{Matthew B. Hastings}
\affiliation{Duke University, Department of Physics, Durham, NC 27708, USA}
\affiliation{Station Q, Microsoft Research, Santa Barbara, CA 93106-6105}
\author{Chetan Nayak}
\affiliation{Station Q, Microsoft Research, Santa Barbara, CA 93106-6105}
\affiliation{Department of Physics, University of California, Santa Barbara, CA 93106}
\author{Zhenghan Wang}
\affiliation{Station Q, Microsoft Research, Santa Barbara, CA 93106-6105}

\date{\today}

\begin{abstract}
We introduce and study a class of anyon models that
are a natural generalization of Ising anyons and Majorana fermion zero modes.
These models combine an Ising anyon sector with a sector
associated with $SO(m)_2$ Chern-Simons theory.
We show how they can arise in a simple scenario for electron
fractionalization and give a complete account of their quasiparticles
types, fusion rules, and braiding.
We show that the image of the braid group is finite for a collection of
$2n$ fundamental quasiparticles and is a proper subgroup of the metaplectic
representation of $Sp(2n-2,\mathbb{F}_m)\ltimes H(2n-2,\mathbb{F}_m)$,
where $Sp(2n-2,\mathbb{F}_m)$ is the symplectic group
over the finite field $\mathbb{F}_m$ and $H(2n-2,\mathbb{F}_m)$
is the extra special group (also called the $(2n-1)$-dimensional Heisenberg group)
over $\mathbb{F}_m$.
Moreover, the braiding of fundamental quasiparticles can be efficiently simulated classically.
However, computing the result of braiding a certain type of composite quasiparticle
is $\# P$-hard, although it is not universal for quantum
computation because it has a finite braid group image.
This a rare example of a topological
phase that is not universal for quantum computation through braiding
but nevertheless has $\# P$-hard link invariants.
We argue that our models are closely related to
recent analyses finding non-Abelian anyonic properties
for defects in quantum Hall systems, generalizing Majorana
zero modes in quasi-1D systems.
\end{abstract}

\maketitle


\section{Introduction}

Majorana zero modes can occur in a wide variety of
physical systems linked by the common thread of
chiral $p$-wave superconductivity and its analogs
\cite{Moore91,Nayak96c,Volovik99,Read00,Cooper01,Kitaev01,Kitaev06a,Lee07,Levin07,Bonderson08a,Nayak08,Fu08,Sato09,Sato10,Sau10a,Lutchyn10,Oreg10,Alicea11,Fidkowski11}.
They exhibit many (and, in some cases, nearly all) of the properties
of Ising anyons and, therefore, may prove useful for
fault-tolerant topological quantum information processing \cite{Kitaev97,Nayak08}.
However, it is possible to classically simulate the braiding
of Ising anyons efficiently \cite{Freedman02a,Freedman02b}.
Therefore, they are useful for quantum
computation only if braiding is supplemented by measurement
at intermediate stages of computations and by a $\pi/8$ phase gate,
in which case they are capable of universal quantum computation \cite{Bravyi01}.
While it is likely that the former can be performed accurately,
the latter appears difficult, although there are various interesting
concrete proposals \cite{Bravyi01,Freedman06,Bonderson10a}.
Moreover, a non-topological implementation of the
$\pi/8$ phase gate requires error correction, which entails significant
overhead \cite{Bravyi06}. Therefore, physical systems supporting anyons
that are capable of universal quantum computation with braiding
alone \cite{Freedman02a,Freedman02b} (best-case scenario) or
braiding and measurement \cite{Mochon03,Mochon04}
(next-best scenario) would be a very attractive platform for quantum computation.

In this paper, we introduce a sequence of topological
phases of electrons which generalize physical
models of Ising anyons. Suppose that an electron
fractionalizes into a spinless neutral fermion $\psi$ and a charged
spinful boson $Z$. Further, suppose that the spinless neutral fermion
forms a $p+ip$ paired superfluid state. If the bosons form
a trivial gapped state, then the system is in the Ising anyon state,
as in Kitaev's honeycomb lattice model \cite{Kitaev06a}.
(If the bosons condense, then the system is in a superconducting state which is a quasi-topological
phase with some of the properties of Ising anyons \cite{Read00,Bonderson12}.)
If the bosons form a spin-polarized
fractional quantum Hall state, then the system is in the Moore-Read state \cite{Moore91},
the anti-Pfaffian state \cite{Lee07,Levin07}, or a Bonderson-Slingerland \cite{Bonderson08a}
state descended from one of these. But suppose, instead, that the bosons form a more complex
topological phase of their own, ${\cal T}$. Then the system will support quasiparticles
that are combinations of those of the Ising topological quantum field theory (TQFT)
and those of ${\cal T}$, subject to the condition that they braid trivially with electrons.
In the phases analyzed in this paper, ${\cal T}$ is associated with $SO(m)_2$
Chern-Simons theory, where $m=3,5,7$ with, we believe, a generalization to
any odd prime $m$. The $SO(m)_2$ TQFTs have several very interesting properties.
All of these theories have a quasiparticle that is a boson. We identify this boson
with $Z$ through a non-Abelian analog of flux-attachment
\cite{Zhang89,Read89a,Jain89,Greiter90,Lopez91}.
In addition, these theories have a `fundamental' quasiparticle,
which we call $X$, that acts as a vortex for the $Z$ boson. $X$ quasiparticles
are non-Abelian anyons with quantum dimension $\sqrt{m}$.
We will call them {\it metaplectic anyons}, for reasons that we will explain.
When two $X$ particles are fused, the result can either be the vacuum
or one of a set of quasiparticles which we call $Y_i$, with $i=1,2,\ldots,r$,
and $r=(m-1)/2$. The $Y_i$ particles have quantum dimension $2$, but this does not mean
that they are trivial; they are also non-Abelian anyons. Finally, there is a particle $X'$,
which results when $X$ and $Z$ are fused. Only a subset of the tensor
product of the quasiparticles of the $SO(m)_2$ TQFT and the quasiparticles
$I,\sigma,\psi$ of the Ising TQFT satisfy the constraint that they braid
trivially with the electron $\Psi_{\rm el} \equiv \psi \cdot Z$, as we will
describe in detail. We call the resulting topological phases
{\it metaplectic-Majorana} TQFTs.

A collection of $N$ quasiparticles of type $X$ at fixed positions
has an $n_N$-dimensional degenerate state space
in the $SO(m)_2$ TQFT with $n_N \sim m^{N/2}$.
Braiding these quasiparticles generates unitary transformations
in $U({n_N})$. These unitary transformations form a finite group,
as in the case of Ising anyons, but unlike Fibonacci anyons.
Therefore, it is not possible to make a universal quantum computer
purely by braiding $X$ particles. We show that braiding can be
efficiently simulated by a classical computer by showing that braiding
operations satisfy a generalization of the Gottesman-Knill theorem \cite{Gottesman98,Nielsen00}.
Indeed, the link invariants computed by these particles in a braiding process
is known to be classically computable in polynomial time.
However, the $Y_i$ particles -- which one might naively
expect to be trivial since they have integer quantum dimensions --
compute a link invariant (the Kauffman polynomial \cite{Kauffman90} at special points)
that is $\#P$-hard \cite{Welsh93}. Therefore, braiding $Y_i$ particles
cannot be efficiently simulated classically. This does not mean that
we can solve $\#P$-hard problems since that would entail
measuring the amplitude for a process with arbitrary
accuracy. Indeed, as we show, the most straightforward approach to encoding quantum
information in $Y_i$ particles leads to a computational model that
can be efficiently simulated classically, and the image of the braid group
of $Y_i$ particles is finite. Nevertheless, the $\#P$-hardness
of braiding $Y_i$ particles hints that metaplectic anyons and
metaplectic-Majorana anyons may have computational
power beyond a classical computer, in spite of the fact that they
cannot serve as a universal quantum compute.
In this respect, they may be similar to the linear optics model of Ref. \onlinecite{Aaronson10}.

We will argue that our topological phase of metaplectic anyons
is closely related to a set of recently proposed two-dimensional
\cite{Barkeshli11b} and quasi-one-dimensional systems
\cite{Clarke12,Lindner12,Cheng12,Vaezi12}.
In these systems, there are defects with interesting topological properties.
In Ref. \onlinecite{Barkeshli11b}, they are dislocations in a fractional
quantum Hall state in a Chern number 2 band.
In Refs. \onlinecite{Clarke12,Lindner12,Cheng12}, the defects live
at the edge of a fractional topological insulator or the edge between two $\nu=1/m$
quantum Hall states that are oppositely spin-polarized. There are two
types of gapped edges, and a defect lives at the point-like boundary
between the two types of gapped edges, generalizing
the $m=1$ case, in which they are Majorana zero modes.
A form of braiding can be defined for the defects in
these models. We show that this braiding operation
is projectively equal to that of $\sigma\cdot X$ quasiparticles in
the metaplectic-Majorana TQFT. However, there are important differences
between metaplectic-Majorana anyons and the defects in these models,
as we will discuss.

We also note that related topological phases have been
constructed in Refs. \onlinecite{Barkeshli10,Barkeshli11a,Barkeshli12a}.
These topological phases have similar anyons with similar
quantum dimensions and topological spins, but it is not clear
what the precise relation is to our phases.

\section{Slave Particle Formulations}

In this section, we give two slave particle descriptions
of electronic systems in the topological phases that
we discuss in the remainder of this paper.
The first is a `parton' model\cite{Wen99} in which the electron
operator is rewritten in terms of partons, each of which
condenses in a simpler topological phase. The second
is a non-Abelian analog of the flux attachment operation that transforms
electrons into `composite bosons' \cite{Zhang89,Read89a} or `composite fermions'
\cite{Jain89,Greiter90,Lopez91}.

For later convenience, we fix notation for $SO(m)$ representations.
We will often write $m$ in the form $m=2r+1$.
We use the standard notation that ${\lambda_1}, {\lambda_2}, \ldots, {\lambda_r}$
are the fundamental weights of $SO(m)$.
The representations with highest weight
${\lambda_1}, {\lambda_2}, {\lambda_3}, \ldots, \lambda_{r-1}, 2{\lambda_r}$
correspond to the representations of $SO(m)$ on, respectively,
vectors; two-index anti-symmetric tensors; three-index anti-symmetric tensors;
\ldots; $(r-1)$-index anti-symmetric tensors; and $r$-index anti-symmetric tensors
(with all indices running from $1$ to $m$).
The representation with highest weight $2{\lambda_1}$ is the representation
of $SO(m)$ on two-index symmetric traceless tensors. The representation
with highest weight ${\lambda_r}$ is the spinor representation of $SO(m)$.

We first consider the following representation of the electron annihilation
operator:
\begin{equation}
\Psi^{\rm el}(x) = f(x)\,C_{\alpha\beta}{\chi^1_\alpha}(x){\chi^2_\beta}(x)
\end{equation}
Here, $f$, ${\chi^1_\alpha}$, and ${\chi^2_\beta}$ are fermions
and $\alpha,\beta = 1,2,\ldots,{2^r}$. $C_{\alpha\beta}$ is the intertwiner
between two copies of the spinor representation of $SO(m)$ and the trivial representation.
This expression for the electron is highly redundant,
as is reflected in its $U(1) \times O(m)$ gauge symmetry.
The $U(1)$ gauge transformation is:
\begin{equation}
f(x) \rightarrow e^{2i\theta}\,f(x) \, , \,\, \chi^{1,2}_{\alpha}(x) \rightarrow 
e^{i\theta}\,\chi^{1,2}_{\alpha}(x)
\end{equation}
while the $O(m)$ gauge transformation is:
\begin{equation}
{\chi^{1,2}_\alpha}(x) \rightarrow O_{\alpha\beta}(x) \,{\chi^{1,2}_\beta}(x)
\end{equation}
We now suppose that the fermions $f$ condense in a $p+ip$ superconducting state
while the fermions $\chi^{1,2}_{\alpha}$ are in gapped insulating states in which they
fill a band with Chern number equal to
$1$. Integrating out the fermions $\chi^{1,2}_{a}$, we generate a Chern-Simons
term at level $2$ for the $SO(m)$ gauge field. (Note that we could, alternatively, consider
a representation of the electron operator in which
$\chi^{1}_{\alpha}=\chi^{2}_{\alpha}$ but these fermions are in a gapped insulating state
in which they fill a band with Chern number equal to $2$.)
Meanwhile, the excitations
of a $p+ip$ superconductor (coupled to a $2+1$-D $U(1)$ gauge field, which
eliminates the Goldstone boson by the Anderson-Higgs mechanism)
are those of the Ising TQFT. Naively, the excitations of this phase are simply
those of $SO(m)_2$ (which we will discuss in detail in the next section) tensored
with those of the Ising TQFT. However, a vortex in the $p+ip$ superconductor
of $f$-pairs will be accompanied with one half of a flux quantum in the
Chern insulating states of $\chi^{1,2}_{a}$. This flux will be produce a
$\chi^{1,2}_{a}$ quasiparticle, carrying the spinor representation of $SO(m)$.
Thus, a $\sigma$ quasiparticle in the Ising sector of the theory is accompanied by
a quasiparticle in the spinor representation of $SO(m)$.

We now consider a (related and, possibly, dual) slave fermion description
of an electron system in which we write the electron annihilation
operator as:
\begin{equation}
\Psi_\alpha^{\rm el}(x) = f(x) \, {z_\alpha}(x)
\end{equation}
where $f$ is a neutral, spinless fermion and $z_\alpha$
is a charge-$e$, spin-$1/2$ boson, and $\alpha=\uparrow,\downarrow$.

We now rewrite the fields ${z_\alpha}$ in terms of auxiliary fields
in a non-Abelian analog of the flux attachment operation that transforms
electrons into `composite bosons' \cite{Zhang89,Read89a} or `composite fermions'
\cite{Jain89,Greiter90,Lopez91}. This is simply a rewriting of the model,
and the original model and the rewritten model
would have the same solution if we could solve them exactly.
However, this re-writing suggests an approximation that we might not
otherwise consider.

We replace the fields ${z_\alpha}$ by 
auxiliary bosons ${Z_\alpha}$ coupled to two
$SO(m)_1$ Chern-Simons gauge fields, $a^1$, $a^2$.
The fields $Z_\alpha$ are $m\times m$ matrices that transform
under $SO(m)\times SO(m)$ as ${Z_\uparrow}\rightarrow {O_2}{Z_\uparrow}{O_1}$
and ${Z_\downarrow}\rightarrow {O_2^T}{Z_\downarrow}{O_1}$,
i.e. they transform in the fundamental representation of both $SO(m)$s.
An $SO(m)_1$ Chern-Simons gauge field would make $Z_\alpha$ into a fermion.
Therefore, two such gauge fields leave $Z_\alpha$ a boson.
In terms of these fields, the Lagrangian then takes the form
\begin{multline}
{\cal L} =
{Z^\dagger_\uparrow}\left(i\partial_0 - {a^1_0} - {a^2_0}\right){Z_\uparrow}
+ \frac{1}{2m_Z}\left| \left(i\partial_0 - {a^1_i} - {a^2_i}\right){Z_\uparrow}\right|^2 \\
+ {Z^\dagger_\downarrow}\left(i\partial_0 - {a^1_0} + {a^2_0}\right){Z_\downarrow}
+ \frac{1}{2m_Z}\left| \left(i\partial_0 - {a^1_i} + {a^2_i}\right){Z_\downarrow}\right|^2\\
+ {f^\dagger}\left(i\partial_0 - {\alpha_0}\right){f}
+ \frac{1}{2m_f}\left| \left(i\partial_0 + {\alpha_i}\right){f}\right|^2\\
+ V({Z_\alpha},f,{f^\dagger}) + {\cal L}_\text{CS}({a_1}) + {\cal L}_\text{CS}({a_2})
\end{multline}
The relation between the original fields ${z_\alpha}$ and the new fields
${Z_\alpha}$ is:
\begin{eqnarray}
{z_\uparrow}(x) &=& {\cal P} e^{-i\int_\infty^x {a_2}} \,\, {Z_\uparrow}(x) \,\,
{\cal P} e^{i\int_\infty^x {a_1}}\cr
{z_\downarrow}(x) &=& {\cal P} e^{i\int_\infty^x {a_2}} \,\,{Z_\downarrow}(x) \,\,
{\cal P} e^{i\int_\infty^x {a_1}}
\end{eqnarray}
We now assume that $Z_\downarrow$ condenses, thereby breaking
$SO(m)\times SO(m)$ to the diagonal $SO(m)$. The Meissner effect
due to $Z_\downarrow$ forces ${a^1_\mu} = {a^2_\mu}$, which we now
write simply as $a_\mu$. The two Chern-Simons terms
then add, and $a_\mu$ has level $2$.

We are now left with $Z_\uparrow$, coupled to an $SO(m)_2$ Chern-Simons
gauge field. Decomposing $Z_\uparrow$ into irreducible representations
of $SO(m)$, we have fields carrying the trivial representation, and the representations
with highest weights $\lambda_2$ and $2\lambda_1$.
Since ${\pi_1}(SO(m)\times SO(m)/SO(m))=\mathbb{Z}_2$, there are also
topological defects in the $Z_\downarrow$ condensate. By forming combinations
of the irreps in $Z_\uparrow$ and the topological defects in
$Z_\downarrow$, we have particles carrying
all of the allowed representations of $SO(m)_2$, namely representations
with highest weights $0, {\lambda_1}, {\lambda_2}, \ldots, {\lambda_r}, 2{\lambda_r},
{\lambda_1}+{\lambda_r}, 2{\lambda_1}$.
We will call the $SO(m)_2$ TQFT the {\it metaplectic TQFT}, for a reason
to be explained when we discuss quasiparticle braiding.

The fermions $f$ are assumed to condense in a $p+ip$ paired state.
Therefore, there are, in addition to the particles listed above,
vortices $\sigma$ and fermions $\psi$. This breaks the $U(1)$
gauge symmetry $f\rightarrow e^{i\theta}f, z\rightarrow e^{-i\theta}z$
down to a $\mathbb{Z}_2$ symmetry. Consequently, $\sigma$ particles,
which are vortices in the $\left\langle ff\right\rangle$ condensate
are accompanied by $\mathbb{Z}_2$ flux which also inserts a topological
defect in the $Z_\downarrow$ condensate. As we will discuss in the next section,
this means that only certain combinations of the particles in the Ising TQFT and
the particles in the metaplectic TQFT are allowed. We dub this combination
the {\it metaplectic-Majorana TQFT}.

\section{Particle Types, Topological Spins, and Fusion Rules}

We introduce the following notation for these quasiparticles.
The particles carrying $SO(m)$ representations
${\lambda_r}$ and ${\lambda_1}+{\lambda_r}$ will be called
$X$ and $X'$. The particles carrying representations
${\lambda_1}, {\lambda_2}, \ldots, \lambda_{r-1}, 2{\lambda_r}$
will be called ${Y_1},{Y_2},\ldots,Y_{r-1},Y_{r}$.
Finally, the particle carrying $2{\lambda_1}$ will be called $Z$.
The particle carrying the trivial representation of $SO(m)$
is equivalent to the vacuum from a topological point of view.
We note that the special case $m=3$ is equivalent
to $SU(2)_4$, and the $X,{Y_1},X',Z$ particles correspond to
spins $\frac{1}{2},1,\frac{3}{2},2$.

The topological properties of the metaplectic TQFT are as follows \cite{Rowell12,Naidu11}.
The topological spins $\theta_a = e^{2\pi h_a}$ of these particles are given by
${h_I} = 0, {h_Z}=1, {h_X}=\frac{r}{8}, h_{X'}=\frac{r+4}{8}, h_{Y_j}=\frac{j(m-j)}{2m}$.
Their fusion rules are:
\begin{eqnarray}
\label{eqn:fusion}
X \cdot X &=& I + {\sum_i} {Y_i}\, , {\hskip 0.5 cm}
X \cdot X' =  Z + {\sum_i} {Y_i}\, , \cr
X \cdot Z &=& X'  \, , {\hskip 1.55 cm}
Z \cdot {Y_i} = Y_i  \, ,\cr
X \cdot {Y_i} &=& X + X' \, , {\hskip 0.8 cm}
Z \cdot Z = I \, ,\cr
Y_i \cdot Y_j &=& Y_{|i-j|} + Y_{\text{min}(i+j,m-i-j)}\, ,\mbox{for $i\neq j$}\cr
Y_i \cdot Y_i &=& I + Z + Y_{\text{min}(2i,m-2i)}
\end{eqnarray}
For the $m=3$ case, there is a single ${Y_i}$, which we will simply call $Y\equiv Y_1$,
and the last of these fusion rules is modified to $Y\cdot Y=I + Z + Y$.
We obtain the dimensions of multi-particle Hilbert spaces from these
fusion rules. If we denote the Hilbert space of $n$ particles of type $X$
with total charge $Q$ by ${\cal H}^Q_{n,X}$, then
\begin{multline}
\label{eqn:multi-qp-dims}
\text{dim}({\cal H}^{I,Z}_{2n,X}) =
 \mbox{$\frac{1}{2}$}(m^{n-1}\pm1)
\, ,
\,\, \text{dim}({\cal H}^{Y_i}_{2n,X}) = m^{n-1}\\
\text{dim}({\cal H}^{X,X'}_{2n+1,X}) =  \mbox{$\frac{1}{2}$}(m^{n}\pm1)\, .
\end{multline}

Combining the Ising (see, e.g. Refs. \onlinecite{Kitaev06a,Nayak08})
and metaplectic TQFTs, we naively have the particle types
$\{I,\sigma,\psi \} \times \{I,X,X',{Y_i},Z\}$.
However, some of these are not local with respect to the electron
operator $\Psi_{\rm el} = \psi \cdot Z$. The topologically-distinct
ones that are local with respect to the electron are: $I, \sigma X, \psi, {Y_i}, Z$.
These $4+r$ particle types determine, for instance, the ground state degeneracy
of the metaplectic-Majorana TQFT on the torus. However, it is worth noting that
this is actually a $\mathbb{Z}_2$-graded TQFT, and one should also consider
as distinct the particle types that differ from these $4+r$ particle types by a single
electron: $\psi Z, \sigma X', Z, \psi {Y_i}$.

Turning now to the particles allowed in the full metaplectic-Majorana
TQFT, we have:
\begin{multline}
\label{eqn:multi-qp-dims2}
\text{dim}({\cal H}^{I,Z\psi}_{2n,\sigma X}) = 2^{n-1}\!\left(\mbox{$\frac{m^{n-1}\pm 1}{2}$}\right),
\,\, \text{dim}({\cal H}^{{Y_i}}_{2n,\sigma X}) = (2m)^{n-1}\\
\text{dim}({\cal H}^{\sigma X,\sigma X'}_{2n+1,X}) =  2^{n-2}(m^{n}\pm1)
\end{multline}

\section{$F$- and $R$-matrices}

We can determine the braiding properties of these particles using their $F$ and $R$
matrices. There are many non-trivial $F$-matrices for $SO(m)_2$, which can be obtained
by solving the pentagon identity. Some, which we will use below,
are\cite{Hong12}:
\begin{eqnarray}
\label{eqn:some-F-matrices}
F^{X Y_1 Y_1}_{X} &=& F^{X' Y_1 Y_1}_{X'} =
\frac{1}{\sqrt{2}}\begin{pmatrix}1 & 1\\ 1& -1 \end{pmatrix}\,, \cr
F^{X Y_1 Y_1}_{X'} &=& F^{X' Y_1 Y_1}_{X} =
\frac{1}{\sqrt{2}}\begin{pmatrix}1 & -1\\ 1& 1 \end{pmatrix}\,,\cr
F^{Y_1 Y_1 Y_1}_{Y_1} &=&
\frac{1}{2}\!\begin{pmatrix}1 & \sqrt{2} & 1 \\ \sqrt{2} & 0 & -\sqrt{2} \\ 1 & -\sqrt{2} & 1
 \end{pmatrix}
\end{eqnarray}
The $F^{X X X}_{X}$-matrix is an $(r+1)\times(r+1)$ matrix.
For $m=3, 5$, it is given by, respectively,\footnote{There are two versions of these theories
which differ by the Froebenius-Schur indicator, which accounts for the minus sign in
these $F$-matrices as well as a few other differences.}
\begin{eqnarray}
F^{X X X}_{X} &=&
-\frac{1}{3}\!\begin{pmatrix} \sqrt{3} & \sqrt{6} \\
\sqrt{6} & -\sqrt{3}
\end{pmatrix}
\cr
F^{X X X}_{X} &=&
-\frac{1}{5}\!\begin{pmatrix} \sqrt{5} & \sqrt{10} & \sqrt{10}\\
\sqrt{10} & -\frac{1}{2}(5+\sqrt{5}) & \frac{1}{2}(5-\sqrt{5}) \\
\sqrt{10} & \frac{1}{2}(5-\sqrt{5}) & -\frac{1}{2}(5+\sqrt{5}) \end{pmatrix}
\end{eqnarray}

Similarly, the $R$-matrices can be obtained by solving the hexagon identity.
Some of the non-trivial ones, which we will use below, are:
\begin{eqnarray}
R^{XX}_{Y_j} &=& i^{(r-j)(r-j+1)-j}\,e^{\pi i(\frac{r}{2}+\frac{j^2}{4r+2})} \, ,\cr
R^{Y_1 Y_1}_{I} &=& e^{\pi i(m+1)/m}\, , \,\,\, R^{Y_1 Y_1}_{Z} = e^{\pi i/m}\, ,\cr
R^{Y_1 Y_1}_{Y_2} &=& e^{\pi i(m-1)/m}\, ,
R^{X Z}_{X'} = i \, , \,\,\, R^{X' Z}_{X} = -i
\end{eqnarray}

With these $F$- and $R$-matrices, we can compute how the
states in the multi-quasiparticle Hilbert spaces of dimensions (\ref{eqn:multi-qp-dims})
transform under braiding.

\section{$N$-particle Braid Group Representations}
\label{sec:braid-group-rep}

We now consider a situation in which we have $n$ particles of type $X$
in the $SO(m)_2$ TQFT. Braiding these particles leads to a representation $\rho_{X}$
of the $n$-particle braid group, ${\mathcal B}_{n}$.
We now describe this representation and its image.
Let $\rho_{X}({\sigma_i})$ be the representative
of the braid group generator $\sigma_i$
(a counter-clockwise exchange of particles $i$ and $i+1$)
acting on the $n$-particle Hilbert space.
From the $R$-matrices, we see that the eigenvalue equation for
$\rho_{X}({\sigma_i})$ is
\begin{equation}
\prod_{j=0}^{r}\left(\rho_{X}({\sigma_i}) - i^{(r-j)(r-j+1)-j}\,e^{\pi i(\frac{r}{2}+\frac{j^2}{4r+2})}\right) = 0
\end{equation}
or, equivalently,
\begin{equation}
\label{eqn:braid-group-eigenvalues}
\prod_{j=0}^{r}\left(i^{r/2}\rho_{X}({\sigma_i}) - i^{-r^2} \,\omega^{j^2}\right) = 0
\end{equation}
where $\omega=e^{2\pi i/m}$.

Consequently, we can represent the braid group in the following way.
We define the {\it extra special group} $H(n,{\mathbb{F}_m})$
(sometimes called the Heisenberg group)
generated by $z,{u_1}, {u_2}, \ldots, u_{n}$,
satisfying the relations
\begin{eqnarray}
\label{eqn:extra-special}
{u_i^m} &=& 1\, ,\,\,{z^m} \,= \, 1 \cr
u_{i} u_{i+1} &=& z u_{i+1} u_{i}\cr
u_{i} u_{j} &=& u_{j} u_{i} \, , {\hskip 0.25 cm} |i-j|>1\cr
u_{i} z &=& z u_{i}
\end{eqnarray}
This is a group of order $m^{n+1}$ which is discussed further in
Appendix \ref{sec:metaplectic}. We introduce this group
because, given a representation of
$H(n,{\mathbb{F}_m})$ by operators ${\hat u}_i$ acting on a
vector space, we can define a representation $\rho_{X}$
of the braid group $B_n$, as we we will see below
and will discuss in further detail in Appendix \ref{sec:metaplectic}.
We construct a representation of $H(n,{\mathbb{F}_m})$
of the requisite dimension as follows. Suppose, for the sake of
concreteness, that $n$ is even and that we are interested in
${\cal H}^{I}_{n}$. Then, we can define
${\cal H}^{I}_{n}=\text{span}(\left| {k_1}, {k_2}, \ldots, k_{n/2}\right\rangle)$
with ${k_i}\in\mathbb{F}_m$,
and define the action of $H(n,{\mathbb{F}_m})$ on ${\cal H}^{I}_{2n}$ by
\begin{multline}
{\hat u}_{2i-1} \left| {k_1}, \ldots, k_{n/2}\right\rangle =
\omega^{2k_{i}} \left| {k_1}, \ldots, k_{n/2}\right\rangle\\
{\hat u}_{2i} \left| {k_1}, \ldots k_{n/2}\right\rangle =
\left| {k_1}, \ldots, {k_i}-1, k_{i+1}+1\ldots, k_{n/2}\right\rangle\\
{\hat z} \left| {k_1},  \ldots, {k_i}, \ldots, {k_n}\right\rangle =
\omega^{-2} \left| {k_1}, \ldots, k_{n/2}\right\rangle
\end{multline}
We could have represented ${\hat z}$ by any $m^{\rm th}$-root
of unity, but we have chosen $\omega^{-2}$ for later convenience.

With this representation of $H(n,{\mathbb{F}_m})$ in hand,
we define a representation $\rho_{X}$ of the braid group $B_n$ according to:
\begin{equation}
\label{eqn:braid-group-rep}
\rho_{X}({\sigma_i}) = \frac{1}{\sqrt{m}}\, i^{-({r^2}+r/2)}\sum_{j=0}^{m-1} \omega^{j^2} {\hat u}_i^j
\end{equation}
Direct computation shows that $\rho_{X}({\sigma_i})$ obeys the Yang-Baxter equation.
Moreover, the states $\sum {\omega^k}|{u_i^k}\rangle$ are eigenvectors
of the braid generator (\ref{eqn:braid-group-rep}) with the
same eigenvalues as Eq. (\ref{eqn:braid-group-eigenvalues})
by virtue of the quadratic Gauss sum,
$\frac{1}{\sqrt{m}}\sum \omega^{j^2} \omega^{jk} = \omega^{-k^2}$.
The eigenvalues and dimensions determine the characters of the representation
which, in turn, determine the representation.
Therefore, we conclude that (\ref{eqn:braid-group-rep}) is the representation
(\ref{eqn:braid-group-eigenvalues}) for $n$ $X$-particles. This representation
of the braid group is called the {\it Gaussian representation} \cite{Jones89}.

We note in passing that there is another possible braid group
representation on this Hilbert space, the {\it Potts representation} \cite{Jones89},
in which $\rho({\sigma_i})= (t+1)\frac{1}{m}\sum_{j=0}^{m-1} u_i^j - 1$,
and $2+t+t^{-1}=m$. The Potts and Gaussian representations coincide
for $m=3$, but differ for $m\geq 5$, where the Potts representation is not
relevant to $SO(m)_2$ since the eigenvalues of the braid group generators
are different. Note that the $m=3$ Potts representation is {\it not} related to
the critical point of the ferromagnetic $3$-state Potts model,
which is the theory of $\mathbb{Z}_3$ parafermions; it is, instead,
related to the critical point of the {\it anti-ferromagnetic} $3$-state Potts model
\cite{Saleur91}.

The image of the braid group in the Gaussian representation
can be understood as follows (see Appendix \ref{sec:metaplectic}
and Refs. \onlinecite{Jones89,Rowell12} for further details).
From Eqs. (\ref{eqn:extra-special})
and (\ref{eqn:braid-group-rep}), we see that
\begin{eqnarray}
\left[\rho_{X}(\sigma_{i+1})\right]^\dagger {u_i}\, \rho_{X}(\sigma_{i+1}) &=& \omega^{-1} u_{i+1} u_i\cr
\left[\rho_{X}(\sigma_{i-1})\right]^\dagger {u_i}\, \rho_{X}(\sigma_{i-1}) &=& \omega u^{-1}_{i-1} u_i\cr
\left[\rho_{X}(\sigma_{i})\right]^\dagger {u_i}\, \rho_{X}(\sigma_{i}) &=& u_i\cr
\left[\rho_{X}(\sigma_{j})\right]^\dagger {u_i}\, \rho_{X}(\sigma_{j}) &=& u_i \, , \,\,|i-j|>1
\end{eqnarray}
Therefore, braiding transforms any $u_i$ into a product of ${u_j}$s,
up to factors of $\omega$. If we mod out
by the factors of $\omega$, then we have $H(n-1,m)/Z(H(n-1,m))$,
the extra special group modulo its center. Braiding transformations
are, therefore, automorphisms of $H(n-1,m)/Z(H(n-1,m))$.
Hence, the image of the braid group is a subgroup
of the group of automorphisms of $H(n-1,{\mathbb{F}_m})/Z(H(n-1,m))$.
As we discuss in Appendix \ref{sec:metaplectic},
this is equal to the {\it metaplectic representation} \cite{Goldschmidt89} of
$Sp(n-1,\mathbb{F}_m)$ for $n$ odd and
$Sp(n-2,\mathbb{F}_m)\ltimes H(n-2,m)$ for $n$ even.
For this reason, we call $X$ particles {\it metaplectic anyons} and we call
$SO(m)_2$ the {\it metaplectic TQFT}.

The group $Sp(n-2,\mathbb{F}_m)\ltimes H(n-2,m)$ is a natural
generalization of the Clifford group. Recall that the {\it Pauli group} is
composed of products of $\pm$ Pauli matrices for
$n/2$ spins; in our notation, it is equal to $H(n,2)$.
The group of automorphisms of the Pauli group
that are trivial on its center is the {\it Clifford group}, and it is equal
to $Sp(n,\mathbb{F}_2)\ltimes{\mathcal P}_{n/2}$.
In other words, braiding metaplectic anyons generates
a subgroup of the analogue of the Clifford group for qudits, with
$\mathbb{F}_2\rightarrow \mathbb{F}_m$.

Turning now to the full metaplectic-Majorana TQFT,
we combine Eq. (\ref{eqn:braid-group-rep}) with
the braid group representation for Ising anyons \cite{Nayak96c}
\begin{equation}
\label{eqn:braid-group-rep-full}
\rho_{\sigma X}({\sigma_i}) = e^{-\frac{\pi i}{8}}\,i^{-({r^2}+r/2)}
\frac{1}{\sqrt{2}} \sum_{k=0}^{1} e^{i\frac{\pi}{2}{k^2}} v_i^k \,
\frac{1}{\sqrt{m}} \sum_{j=0}^{m-1} \omega^{j^2} u_i^j
\end{equation}
where ${v_i^2}=1$, $v_{i} v_{i+1} = - v_{i+1} v_{i}$,
$v_{i} v_{j} = v_{j} v_{i}$ for $|i-j|>1$.

\section{Quantum Information Processing with the Metaplectic-Majorana TQFT}
\label{sec:quantum-comp}

We will consider three different encodings of quantum information
into the many-particle states of the Metaplectic-Majorana TQFT.
For reasons that will become clear, we call them the `qudit', `qubit',
and `qutrit' encodings.

Consider the state space of $4$ $\sigma X$-particles
with total topological charge $Y_1$. It can be depicted graphically as follows.
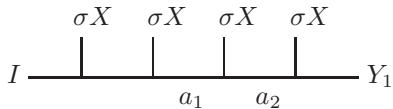
\begin{center}
\begin{picture}(250,30)
\put(62,8){$I$}
\put(87,30){$\sigma X$}
\put(115,30){$\sigma X$}
\put(142,30){$\sigma X$}
\put(169,30){$\sigma X$}
\put(127,0){$a_1$}
\put(156,0){$a_2$}
\put(70,10){\line(1,0){125}}
\put(90,10){\line(0,1){15}}
\put(117,10){\line(0,1){15}}
\put(144,10){\line(0,1){15}}
\put(171,10){\line(0,1){15}}
\put(198,8){$Y_1$}
\end{picture}
\end{center}
The first two particles fuse to $a_1$, which can be
$I, {Y_1}, \ldots {Y_r}, \psi, \psi {Y_1}, \ldots, \psi{Y_r}$.
In all of these cases, ${a_2}=\sigma X$ is possible.
However, if ${a_1}={Y_1}, \ldots {Y_r}, \psi {Y_1}, \ldots, \psi{Y_r}$,
then ${a_2}=\sigma X'$ is also possible.
Therefore, there are $2(r+1)+2r=2m$ such states.
We will take a basis $\left| j,{n_\psi}\right\rangle$ with
$0\leq j < m$ and $n_\psi = 0,1$ for this $2m$-state {\it qudit}.
$\left| j,0\right\rangle$ corresponds, for $0\leq j \leq r$,
to the state with ${a_1}={Y_j}$, ${a_2}={\sigma X}$ (with the notation ${Y_0}\equiv I$)
and, for $r\leq j \leq m-1$, to the state with ${a_1}=Y_{m-j}$,
${a_2}={\sigma X'}$. Meanwhile, $\left| j,1\right\rangle$ corresponds, for $0\leq j \leq r$,
to the state with ${a_1}=\psi {Y_j}$, ${a_2}={\sigma X}$ (with the notation ${Y_0}\equiv I$)
and, for $r\leq j \leq m-1$, to the state with ${a_1}=\psi Y_{m-j}$,
${a_2}={\sigma X'}$.

For such a qudit, there are two
generators of the unitary transformations that can be performed by braiding.
The first is a counter-clockwise exchange of the two $\sigma X$-particles
on the left. This implements the following gate which is diagonal in this basis:
\begin{equation}
\rho({\sigma_1})\left| j,{n_\psi}\right\rangle =\\
e^{-\frac{\pi i}{8}} i^{r(r+2)} e^{\frac{\pi i}{2}{n^2_\psi}}\omega^{j^2}
\left| j,{n_\psi} \right\rangle
\end{equation}
The second is a counter-clockwise exchange of the middle two
$\sigma X$-particles. This can be obtained by using the $F$-matrix to
transform into a basis in which these two particles have a fixed fusion channel,
applying the $R$ matrix, and transforming back, i.e. from $F^{-1} R F$.
For the sake of concreteness, let us consider the case $m=3$.
Then $\rho({\sigma_2})\left| j,{n_\psi}\right\rangle = M_{jk} L_{{n_\psi}{n'_\psi}}
\left| k,{n'_\psi}\right\rangle$ where
\begin{equation}
M = \begin{pmatrix}
\mbox{$\frac{1}{3}$}(1+2\omega) & \mbox{$\frac{\sqrt{2}}{3}$}(1-\omega) & 0\\
\mbox{$\frac{\sqrt{2}}{3}$}(1-\omega) &  \mbox{$\frac{1}{3}$}(2+\omega) & 0\\
0 & 0 & \omega
\end{pmatrix}\!, \,
L = \frac{e^{\frac{\pi i}{8}}}{\sqrt{2}}\begin{pmatrix}
1 & -i\\
-i & 1
\end{pmatrix}
\end{equation}
In a similar manner, we obtain the gate associated with a
a counter-clockwise exchange of the last two
$\sigma X$-particles for $m=3$, which takes the form
$\rho({\sigma_3})\left| j,{n_\psi}\right\rangle = {\tilde M}_{jk}
e^{-\frac{\pi i}{8}}e^{\frac{\pi i}{2}{n^2_\psi}} \left| k,{n_\psi}\right\rangle$
with
\begin{equation}
{\tilde M} = \begin{pmatrix}
\omega & 0 & 0\\
0 & (1+\omega)/2 & (1-\omega)/2 \\
0 & (1-\omega)/2  &  (1+\omega)/2
\end{pmatrix}
\end{equation}

For multiple qudits, we can employ either a dense or sparse encoding.
A dense encoding using $2k$ $\sigma X$-particles can be represented by:
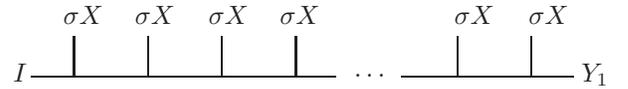
\begin{center}
\begin{picture}(250,30)
\put(3,8){$I$}
\put(22,30){$\sigma X$}
\put(49,30){$\sigma X$}
\put(77,30){$\sigma X$}
\put(104,30){$\sigma X$}
\put(10,10){\line(1,0){115}}
\put(26,10){\line(0,1){15}}
\put(54,10){\line(0,1){15}}
\put(82,10){\line(0,1){15}}
\put(110,10){\line(0,1){15}}
\put(132,10){$\dots$}
\put(150,10){\line(1,0){65}}
\put(171,10){\line(0,1){15}}
\put(199,10){\line(0,1){15}}
\put(170,30){$\sigma X$}
\put(198,30){$\sigma X$}
\put(218,8){$Y_1$}
\end{picture}
\end{center}
Such an encoding uses $2k$ $\sigma X$-particles for $k-1$ qudits.
However, an exchange of neighboring particles will necessarily involve neighboring qubits.
Consequently, simple single-qudit gates are complicated in terms of braids and
errors in one qubit tend to infect others. We can, alternatively,
use a sparse encoding, such as:
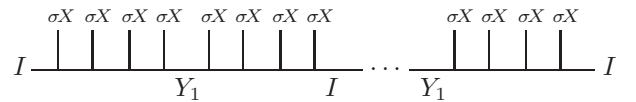
\begin{center}
\begin{picture}(250,30)
\put(3,8){$I$}
\put(16,28){$\scriptstyle{\sigma\!X}$}
\put(30,28){$\scriptstyle{\sigma\!X}$}
\put(44,28){$\scriptstyle{\sigma\!X}$}
\put(57,28){$\scriptstyle{\sigma\!X}$}
\put(10,10){\line(1,0){125}}
\put(20,10){\line(0,1){15}}
\put(33,10){\line(0,1){15}}
\put(47,10){\line(0,1){15}}
\put(60,10){\line(0,1){15}}
\put(64,0){$Y_1$}
\put(74,28){$\scriptstyle{\sigma\!X}$}
\put(87,28){$\scriptstyle{\sigma\!X}$}
\put(101,28){$\scriptstyle{\sigma\!X}$}
\put(114,28){$\scriptstyle{\sigma\!X}$}
\put(77,10){\line(0,1){15}}
\put(90,10){\line(0,1){15}}
\put(104,10){\line(0,1){15}}
\put(117,10){\line(0,1){15}}
\put(121,0){$I$}
\put(138,10){$\dots$}
\put(157,0){$Y_1$}
\put(167,28){$\scriptstyle{\sigma\!X}$}
\put(180,28){$\scriptstyle{\sigma\!X}$}
\put(194,28){$\scriptstyle{\sigma\!X}$}
\put(207,28){$\scriptstyle{\sigma\!X}$}
\put(153,10){\line(1,0){70}}
\put(170,10){\line(0,1){15}}
\put(183,10){\line(0,1){15}}
\put(197,10){\line(0,1){15}}
\put(210,10){\line(0,1){15}}
\put(226,8){$I$}
\end{picture}
\end{center}
In such an encoding, $4k$ $\sigma X$-particles are used
for $k$ qubits. There are $k$ sets of $4$ $\sigma X$-particles.
Each set of $4$ has total topological charge $Y_1$.
These sets of $4$ are paired so that each pair of sets (i.e. a group
of eight $\sigma X$-particles) has total topological charge $I$.

An alternative encoding scheme, which we call the qubit encoding,
uses a $\sigma X$ particle and $(n+1)$ $Y_1$-particles (or any other
$Y_i$) to encode $n$ qubits. It is depicted as follows:
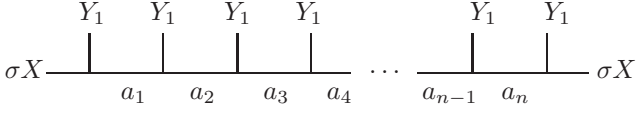
\begin{center}
\begin{picture}(250,30)
\label{pic:qubit-encoding}
\put(2,8){$\sigma X$}
\put(30,30){$Y_1$}
\put(57,30){$Y_1$}
\put(85,30){$Y_1$}
\put(112,30){$Y_1$}
\put(46,0){$a_1$}
\put(72,0){$a_2$}
\put(100,0){$a_3$}
\put(124,0){$a_4$}
\put(160,0){$a_{n-1}$}
\put(190,0){$a_{n}$}
\put(18,10){\line(1,0){115}}
\put(34,10){\line(0,1){15}}
\put(62,10){\line(0,1){15}}
\put(90,10){\line(0,1){15}}
\put(118,10){\line(0,1){15}}
\put(140,10){$\dots$}
\put(158,10){\line(1,0){65}}
\put(179,10){\line(0,1){15}}
\put(207,10){\line(0,1){15}}
\put(178,30){$Y_1$}
\put(206,30){$Y_1$}
\put(226,8){$\sigma X$}
\end{picture}
\end{center}
where ${a_i}=\sigma X$ or $\sigma X'$.
In order to express the gate that results when particles
$i$ and $i+1$ are exchanged, it is useful to define
${H_i}\equiv{X_i}$ if $m=3$ and ${H_i}\equiv Z_{i-1}{X_i}Z_{i+1}$
if $m\geq 5$ (note that $X_i$, $Z_i$ are the usual Pauli matrices
here because we have qubits rather than qudits).
We label the qubits by $i=1,...,n$, and we define $Z_0=Z_{n+1}=+1$.
In addition,
we define $\text{NOT}_i^{-+} \equiv I$ if $Z_{i-1} Z_{i+1}=1$ and
$\text{NOT}_i^{-+} \equiv X_i$ if $Z_{i-1} Z_{i+1}=-1$. Then,
a counter-clockwise exchange of particles $i$ and $i+1$ results in a gate
that can be written in the following form:
\begin{equation}
\rho_{Y_1}({\sigma_i}) = e^{\frac{\pi i}{m}{H_i}}\,\text{NOT}_i^{-+}
\end{equation}

Finally, we introduce one more encoding: the qutrit
representation. (Qutrits are obtained for any $m$.
Note that the qudit representation introduced earlier is never a qutrit representation
since the dimension $2m$ is always even). A qutrit is encoded
in four $Y_1$ particles with total charge $I$:
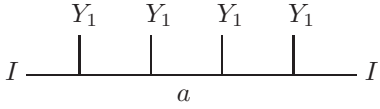
\begin{center}
\begin{picture}(250,30)
\put(62,8){$I$}
\put(87,30){$Y_1$}
\put(115,30){$Y_1$}
\put(142,30){$Y_1$}
\put(169,30){$Y_1$}
\put(127,0){$a$}
\put(70,10){\line(1,0){125}}
\put(90,10){\line(0,1){15}}
\put(117,10){\line(0,1){15}}
\put(144,10){\line(0,1){15}}
\put(171,10){\line(0,1){15}}
\put(198,8){$I$}
\end{picture}
\end{center}
From the fusion rules (\ref{eqn:fusion}), we see that the charge
$a=I,{Y_2},Z$ (except in the case $m=3$, where $a=I,{Y_1},Z$).
Braiding the first two particles enacts the transformation:
\begin{equation}
\rho_{Y_1}({\sigma_1}) = -e^{\pi i/m}\begin{pmatrix}
1 & 0 & 0\\
0 & \overline{\omega} & 0 \\
0 & 0  &  -1
\end{pmatrix}
\end{equation}
while braiding the second two enacts:
\begin{equation}
\rho_{Y_1}({\sigma_2}) = \begin{pmatrix}
2\overline{\omega} & 2\sqrt{2} & -2\overline{\omega}\\
2\sqrt{2} & 0 & 2\sqrt{2} \\
-2\overline{\omega} & 2\sqrt{2}  &  2\overline{\omega}
\end{pmatrix}
\end{equation}

\section{Classical Simulation of Braiding in the Metaplectic-Majorana TQFT}

Regardless of the encoding, universal quantum computation
is not possible purely through braiding because the braid group representation
(\ref{eqn:braid-group-rep}) for $n$ $X$-particles
is contained within $Sp(n-1,\mathbb{F}_m)$ for $n$ odd and
$Sp(n-2,\mathbb{F}_m)\ltimes H(n-2,{\mathbb{F}_m})$ for $n$ even.
 As we discuss in greater detail in Appendix \ref{sec:metaplectic},
\begin{equation}
\left| Sp(2n,\mathbb{F}_m) \right| = m^{n^2} \prod_{i=1}^{n} \left(m^{2i}-1\right)
\end{equation}
while $\left|H(n,m)\right|=m^{n+1}$, so the braid group has a finite image
under the Gaussian representation. Therefore, it is not possible to approximate an
arbitrary unitary transformation to any desired accuracy.
In fact, braiding $\sigma X$ particles can be efficiently simulated by a classical
computer.

Since it is known that braiding in the Ising TQFT can be efficiently simulated classically,
we focus on the braiding of metaplectic anyons.
Recall that braiding metaplectic anyons transforms products of $u_i$s
into products of $u_i$s, as we noted in Section \ref{sec:braid-group-rep}.
As a result, the evolution of eigenstates of such products can be efficiently simulated classically
by following the evolution of these operators. In order to see this in greater detail,
it is convenient to embed $H(n-2,{\mathbb{F}_m})$ inside $H(2n,{\mathbb{F}_m})$
as follows. Let ${X_1},\ldots,{X_n},{Z_1},\ldots,{Z_n},\omega$ be a set of generators of
$H(2n,{\mathbb{F}_m})$, as described in Appendix \ref{sec:metaplectic}
(see, especially, Eq. \ref{eqn:Heisenberg-Z-X}).
Then ${U_i}={X_i}X_{i+1}{Z_i}Z_{i+1}^\dagger$ faithfully represents
the extra special group (\ref{eqn:extra-special}).
Consequently, ${\rho_X}({\sigma_i})=\frac{1}{\sqrt{m}} \sum_{j=0}^{m-1} \omega^{j^2} U_i^j$
represents the braid group.
We can prepare states that are eigenstates of $U_i$
by creating pairs out of the vacuum. Such states are
stabilized by products of ${X_i}$ and ${Z_j}$ operators
since $U_i$ can be expressed as such a product.
To see how any state stabilized by products of ${X_i}$ and ${Z_j}$
operators transforms under braiding, we
can follow the evolution of the operators ${X_i}$, ${Z_j}$. It is sufficient
to consider the case of two qudits. We would like to see how ${X_1},{X_2},{Z_1},{Z_2}$
(and, therefore, the group that they generate)
evolve under the action of $R$. First, note
that we can replace the set ${X_1},{X_2},{Z_1},{Z_2}$ by the set
${Z_1},{X_1}{X_2},{Z_1}{Z_2^\dagger},{X_1}{Z_1}$, which generates the
same group. The latter three commute with $U$ and, therefore, with $R$.
Therefore, we need only study how $Z_1$ evolves.
Using $U^j Z_1 = \omega^{-j} Z_1 U^j$, we see by
direct computation that ${\rho_X}({\sigma_1}) Z_1 {\rho_X^\dagger}({\sigma_1})
= \omega^{-k^2} Z_1 U^k$,
where $k=(m+1)/2$. Therefore, the evolution of $Z_1$ can be
efficiently simulated classically and, as a consequence, so can the evolution
of any state stabilized by products of ${X_i}$ and ${Z_j}$ operators.
Thus, we conclude that we can efficiently simulate classically
any operation that consists of creating pairs of
$X$ particles out of the vacuum, braiding them, and then measuring them
a basis of products of ${X_i}$ and ${Z_j}$ operators (e.g. the $U_{2i-1}$
basis).

Of course, as noted above, $H(2n,{\mathbb{F}_m})$ is much too large.
It associates an $m$-state qudit to each $X$-particle while,
in the dense encoding, there should be a qudit associated to each {\it pair} of $X$-particles.
Therefore, braiding should commute trivially with roughly half of the generators
of $H(2n,{\mathbb{F}_m})$. This is, indeed, the case, as may be seen
by considering the following set of generators of $H(2n,{\mathbb{F}_m})$:
${U_1}, \ldots U_{n-1}, {\tilde U_1}, \ldots {\tilde U}_{n-1}, {X_1}{Z_1},
{X_n}{Z^\dagger_n},\omega$ where ${\tilde U_i}={X_i}X_{i+1}{Z_i}^\dagger Z_{i+1}$.
The generators ${\tilde U_i}$s, ${X_1}{Z_1}$, and ${X_n}{Z^\dagger_n}$ all
commute with the ${U_i}$s and, therefore, with braiding.

Braiding is not universal in the qubit representations, either.
We now show that the group generated by the $\rho_{Y_1}({\sigma_i})$
operators acting on the qubit representation is finite, and we give an efficient classical way to store an arbitrary element of this group and to efficiently compute products of elements of this group with braid generators (the method we describe will only store an element up to an overall phase).
For all $m$ (including both $m=3$ and $m>3$), a direct computation gives
\begin{eqnarray}
\label{conjugate}
\Bigl(\NOT_i\Bigr)^\dagger H_i \NOT_i &=& H_i, \\ \nonumber
\Bigl(\NOT_i\Bigr)^\dagger H_{i+1} \NOT_i &=& H_i H_{i+1}, \\ \nonumber
\Bigl(\NOT_i\Bigr)^\dagger H_{i-1} \NOT_i &=& H_{i-1} H_i, \\ \nonumber
\end{eqnarray}
so conjugating a product of the $H_i$ by a unitary $\NOT_j$ gives some, possibly different, product of the $H_i$.
The group generated by the operators $e^{\frac{\pi i}{m}{H_i}}$ is an Abelian group, which we call $G$.  Since $e^{2\pi H_i}=1$, we can write an arbitrary element of the group as $e^{i \sum_i k_i \frac{\pi}{m}{H_i}}$, where the $k_i$ are integers ranging from $0,...,2m-1$, so the group is a subgroup of $\mathbb{Z}_{2m}^n$.  However, since $e^{\pi H_i}=-1$, there are only $2\cdot m^n$ distinct group elements which can be written as $(\pm 1)\cdot e^{i\sum_i k_i \frac{\pi}{m}{H_i}}$, where the $k_i$ are integers ranging from $0,...,m-1$. This group is in fact $\mathbb{Z}_m^n \times \mathbb{Z}_2$, and the generators of the group can be taken to be
$-e^{i \frac{\pi}{m} H_i}$ and $-1$. The group generated by the operators $\NOT_i$ is a subgroup of the Clifford group; call this group $H$.  Then, because conjugation by
$\NOT_i$ defines an automorphism of $G$, the group generated by
 $e^{i\frac{\pi}{m}{H_i}}\,\text{NOT}_i^{-+}$ is the semi-direct product $G \rtimes H$.  This gives us an efficient way to store elements of the group by storing a list of integers $k_i$ and also storing an element of the Clifford group.  We specify an element $U$ of the Clifford group by specifying $U X_i U^\dagger$ and $U Z_i U^\dagger$ for all $i$.  These products $U X_i U^\dagger$ and $U Z_i U^\dagger$ are products of Pauli matrices and so can be stored efficiently (we are essentially using the Gottesman-Knill theorem here).  Storing these products fully specifies $U O U^\dagger$ for any operator $O$ and so specifies $U$ up to a phase.  To take a product of two elements of the group, say the first being represented by a product $AU$ and the second by a product $A' U'$ where $A,A'$ are in the Abelian group and $U,U'$ are in the Clifford group, we write $A U A' U'=A (U A' U^\dagger) U U'$.  We then compute $U A' U^\dagger$ using our known values of $U X_i U^\dagger$ and $U Z_i U^\dagger$ and the result will be some other element of the Abelian group, all it $A''$.  Then the desired product is $A A'' U U'$, and the product of the first two is in the Abelian group and the product of the second two is in the Clifford group.

It should not be surprising that the group image is finite.  The $Y_1$ particles can be obtained by fusing a pair of $X$ particles.  Thus, the fusion tree in Section \ref{sec:quantum-comp}
that defined the qubit representation can be written as a tree with $2(n+1)+2$ $X$-particles,
with $2(n+1)$ of the $X$-particles fusing in pairs to make $(n+1)$ $Y_1$-particles.  Braiding two $Y_1$ -particles can be done by braiding two pairs of $X$-particles.
Since the image for braiding $X$-particles is finite, it is no surprise that the image for
braiding $Y_1$-particles is also finite.  However, it is still important to check, as we have done,
that we can efficiently store elements of this group; after all, the tree that we have written
here with $2(n+1)+2$ $X$-particles is related by some sequence of $F$ moves to the
previous tree in terms of $X$-particles and it is not immediately obvious that all these
$F$ moves can be computed efficiently.

\section{Computational Complexity of Link Invariants}

In the previous section, we have seen that braiding is
not universal for quantum computation in any representation.
Moreover, braiding in the qudit and qubit representations
can be efficiently simulated classically. However, this theory displays a surprise
when we turn to the computation of link invariants. Thus far, the most-studied examples
of TQFTs for which braiding is universal for quantum computing
have been precisely those for which an evaluation of the link invariants is $\#P$-hard.
However, there seems to be no deep reason why this should be true generally,
and indeed the present theory is not universal for quantum computing
(through braiding alone), but it does have a link invariant
that is $\#P$-hard to compute. Said differently, there are experiments
whose results are $\#P$-hard to predict,
i.e. cannot be predicted with a classical computer
(unless the hierarchy of complexity classes collapses),
even though braiding alone is not sufficient for universal
quantum computation.

We give a more precise definition of this link invariant elsewhere \cite{math-version}.
Here we will give its physical motivation. We imagine creating a collection
of pairs of $Y_1$ particles out of the vacuum. We braid them with each other
and then fuse them again in pairs. There will be some amplitude $E(L)$ for
all of these fusion processes to give the vacuum, i.e. to be annihilation
processes. (When two $Y_1$ particles are fused, the result could be the vacuum $I$,
but it could, instead, be $Z$ or $Y_2$, except in the $m=3$ case, in which there is no
$Y_2$ particle and it could, instead, be $Y_1$.)
Here, $L$ is the link formed by the spacetime trajectories of the $Y_1$ particles.
The amplitude $E(L)$ is our `link invariant'. We use quotation
marks because this amplitude is not necessarily a topological invariant
unless further conditions are satisfied. However,
if the interaction between the $Y_1$ particles decays exponentially (or faster),
then, in the limit that the particles stay far apart while braiding, this amplitude will depend only
on the topological class of the $Y_1$ trajectories. When the particles
are being pair-created and annihilated, the amplitude will acquire a non-topological,
non-universal phase. However, this can be made to cancel between
creation and annihilation. Alternatively, if two different braiding processes
are interfered, then this non-topological phase will cancel.
See, for instance, Refs. \onlinecite{Nayak08,Bonderson08c}
for a discussion of interference measurements for link invariants.

The starting point for the $\#P$-hardness of $E(L)$
is a result of Lickorish and Millett\cite{Lickorish88}.
They show that the link invariant $E(L)$ can be written as
\be
\label{LMeq}
E(L)=\sum_{S \subset L} a^{-4 \langle S,L-S\rangle},
\ee
where
\be
a=-i\exp(-i \pi/m).
\ee
Here, the sum is over links $S$ which are a sublink of link $L$.
A link may be made of more than one disconnected component,
where each component of the link is some knot; we use $c(L)$ to write the
number of components of $L$.  A sublink $S$ contains some subset of the components,
so there is a total of $2^{c(L)}$ terms in the sum, with each factor of $2$ coming from
the choice of whether a given component is in a sublink or not.
We can specify a sublink $S$ by a vector $s$ with entries $s_i$ for $i=1,...,c(L)$,
such that $s_i=+1$ if the $i$-th component is in $S$ and $s_i=-1$ otherwise.
The invariant  $\langle S,L-S\rangle$ is defined to be the sum of $\langle i,j\rangle$ over pairs $i\in S$ and $j\in L-S$, where $\langle i,j \rangle$ is the {\it linking number}
between the $i$-th sublink and the $j$-th sublink.\footnote{The linking number
of two closed oriented curves may be computed by drawing a projection of the link
on a plane, with care taken to denote over-crossings and under-crossings.
Those crossings in which the overcrossing curve goes to the right of the intersection
are called positive. Those in which the overcrossing curve goes to the left of the
intersection are called negative. The linking number is one-half
the number of positive crossings minus the number of negative crossings.
The linking number is symmetric, so that $\langle i,j\rangle=\langle j,i\rangle$.}

Eq.~(\ref{LMeq}) looks very much like the partition function of an Ising model at an imaginary temperature.  The sum over sublinks corresponds to a sum over the ``Ising spin" degrees of freedom $s_i$, while the term $ a^{-4 \langle S,L-S\rangle}$ looks like a complex Boltzmann weight.
To see this, write
\ba
-4 \langle S,L-S\rangle &=& -\sum_{i \neq j} (1+s_i) (1-s_j) \langle i,j \rangle\cr
&=&-2\sum_{i < j} (1-s_i s_j) \langle i,j \rangle
\ea
Consequently Eq.~(\ref{LMeq}) is equal to
\be
\label{boltzmann}
E(L)=a^{-2 \sum_{i < j} \langle i,j \rangle}\sum_{s \in \{-1,1\}^{c(L)}} a^{2 \sum_{i < j} s_i s_j \langle i,j \rangle}.
\ee
So, up to the prefactor in front, the resulting link invariant is the partition function of an Ising spin system with Boltzmann weights
\be
\exp(\beta \sum_{i<j} \langle i,j \rangle  s_i s_j),
\ee
where $\beta=-2\pi i/m+\pi i$.

Note that the temperature is purely imaginary.  The quantity $\langle i,j \rangle$ plays the role of a matrix of coupling constants; note that these linking numbers $\langle i,j \rangle$ can be taken to have any integer values.  In particular, the Ising model need not be planar, and any choice of $\langle i,j \rangle$ can be realized by some link $L$ in which the number of crossings
is at most polynomial in $\sum_{ij} |\langle i,j \rangle|$.

We will now show that there is a class of links for which
we can relate this Ising model with complex Boltzmann weights
to more familiar models with real or even real and positive Boltzmann weights.
We then argue that computing the resulting partition function is
$\#P$-hard. (While we cannot relate $E(L)$ for an arbitrary link
to an Ising model with real Boltzmann weights, it is sufficient to do
so for the class of links discussed below. We can then conclude that if we can
compute $E(L)$ for an arbitrary link, then we can solve any
problem in $\#P$.)

To obtain an Ising model with real or even real and positive
Boltzmann weights, we use the following trick.
We consider links $L$ constructed as follows.
We begin with a link $L'$ with $c(L')=N$ unlinked
components, i.e. for any $i,j\in \{1,2,\ldots,N\}$, $\left\langle i,j \right\rangle =0$.
We then add components $N+1, N+2, \ldots, c(L)$ to form the link $L$.
They are chosen so that if $i,j\in \{1,2,\ldots,N\}$, then
$\left\langle i,k \right\rangle = \left\langle j,k \right\rangle$
(if $k\in \{1,2,\ldots,N\}$, then both sides of the equality are zero,
but if $k\in \{N+1,N+2,\ldots,c(L)\}$, then they might be non-zero).
We now evaluate the link invariant $E(L)$ in two steps.
First, we sum over the choices of $s_k$ for
$k=N+1,\ldots,c(L)$ to define an ``effective Boltzmann weight" for the first
$N$ Ising spin variables. Summing over component $k$ generates an effective interaction between $i$ and $j$ if $\left\langle i,k \right\rangle = \left\langle j,k \right\rangle\neq 0$.
The effective Boltzmann weight will be real and $E(L)$ is equal to the sum
over the $2^N$ choices of the first $N$ spin variables using the effective Boltzmann weigh

Consider a pair $i,j$ with $1\leq i<j \leq N$.  We now add a component $k\in \{N+1,N+2,\ldots\}$,
such that $\langle i,k \rangle= \langle j,k \rangle =\pow$ for some $\pow$ and such that
$\langle k,l \rangle=0$ for $l$ different from $i$ or $j$.  Then, summing over ${s_k}=\pm 1$
will produce an effective interaction between $s_i$ and $s_j$.
Summing over ${s_k}=\pm 1$ gives a weight
\begin{eqnarray}
\label{effW}
 \sum_{s_k \in \{-1,1\}} a^{2s_i s_k \langle i,k\rangle
 +2s_j s_k \langle j,k\rangle} &=&\sum_{s_k \in \{-1,1\}} a^{2 \pow (s_i+s_j) s_k }
\cr
&=& (\sqrt{y})^{s_i s_j} \sqrt{z},
\end{eqnarray}
where
\be
\label{ydef}
y=\frac{a^{-4\pow}+a^{4\pow}}{2},
\ee
and
\be
z=2(a^{-4\pow}+a^{4\pow}).
\ee
and any ambiguity in the sign of the square-root is resolved by choosing $\sqrt{y} \sqrt{z}=a^{-4\pow}+a^{+4\pow}$.

Ignoring the overall factor $\sqrt{z}$, the effective weight is
$(\sqrt{y})^{s_i s_j}$.
By adding additional components $k$, $k'$ of the link and summing over
$s_{k'}, s_{k''}$, and so on, we can replace this weight with any power, so that the effective weight for the first $N$ variables can be chosen to be (again up to an overall factor)
\be
\prod_{1\leq i<j\leq N} (\sqrt{y})^{s_i s_j J_{ij}},
\ee
for any matrix $J_{ij}$ with non-negative integer entries (in fact, it is also possible to obtain negative entries by a slightly different trick but we will not need that here).  The size of the link needed to produce this effective weight is at most  polynomial in $\sum_{ij} |J_{ij}|$.

The quantities $y$ are real.  However, depending upon $m$ and $\pow$, they may be positive or negative.  In fact, for any odd $m>1$, we can choose $-1<y<0$ by an appropriate choice of $\pow$, and for odd $m>3$ we can instead choose $0<y<1$ by an appropriate choice of $\pow$.
One way to obtain positive weights for $m=3$ is to pick the entries of $J_{ij}$ to be even integers.
In this way, we succeed in constructing a link invariant that equals, up to multiplication by a trivial overall constant, the partition function of an Ising model at real, positive temperature with antiferromagnetic couplings.  By taking these couplings large, we can ensure that ground states provide the dominant contribution to the partition function.  That is, that the partition function is equal to $N_0 \exp(-\beta E_0)$ plus a small correction (small compared to $\exp(-\beta E_0)$), where $\beta$ is now real and positive and where $E_0$ is the ground state energy and $N_0$ is the number of ground states.
Making the correction small compared to $\exp(-\beta E_0)$ requires only polyomially large coupling constants (we are choosing the coupling constants large enough that energy outweighs entropy and so the sum of the weights of all the higher energy states is small compared to the weight of a single ground state).  Then, an evaluation of the partition function lets one determine both the ground state energy and also the number of ground states. Counting the number of ground states
is equivalent to finding the number of maximum cuts in a graph which
is a $\#P$-hard problem\footnote{The fact that this problem is $\#P$-hard is a textbook exercise.  See C. Moore and S. Mertens, {\it The Nature of Computation}, Ex. 13.11.}.
Indeed the definition of $\#P$ is that it is the problem of counting the number of solutions to a decision problem in $NP$.

This approach shows that evaluation of the link invariant to exponential accuracy is $\#P$-hard.  In fact, it is possible also to consider the case with negative and real Boltzmann weights (the case $y<0$ but $J_{ij}$ has odd entries).  Then, even the evaluation of the sign of the partition function is $\#P$-hard, as follows from a result of Goldberg and Jerrum\cite{Goldberg12}.  The sign of the partition function is equal to the phase of the link invariant multiplied by some overall phase which can be computed trivially.

Similar behavior is seen in the theory of
\cite{Krovi12}, which also has a finite braid group image but $\#P$-complete
link invariants. It would be interesting to see if
our theory follows the pattern of their theory, where different
approximations to the link invariant are in P, or are SBP-hard, or are
$\#P$-hard, depending upon the accuracy of the approximation.  It would
be interesting to see if their theory, like ours, is classically
simulable for certain measurements

\section{Relation to Fractional Quantum Hall Devices}

In Refs. \onlinecite{Clarke12,Lindner12,Cheng12} (see also Ref.~\onlinecite{Vaezi12}),
a model was presented in which the boundary of
a fractional topological insulator was divided into $2N$
intervals, with the $i^{\rm th}$ interval lying between points $x_{i-1}$
and $x_{i}$ and $x_{0}\equiv x_{2N}$. The even intervals $(x_{2j-1},x_{2j})$
are brought into contact with $s$-wave superconductors, while the
odd intervals $(x_{2j},x_{2j+1})$ are brought into contact with ferromagnets.
The points $x_i$ are viewed as particles. They `fuse' to the $2m$
possible allowed total spins (modulo 1) on the even intervals or
$2m$ possible allowed charges (module 2e) on the odd intervals.
They can be `braided' \cite{Clarke12,Lindner12}
by a measurement-only process \cite{Bonderson08b,Bonderson08d}.
The resulting unitary transformation for braiding two neighboring defects
at $x_k$, $x_{k+1}$ is \cite{Clarke12,Lindner12}:
\begin{equation}
U_{k,k+1} = e^{i\pi q^2 /2m}
\end{equation}
where $q=0,1,\ldots, 2m-1$ are the possible charges/spins on the
interval between the two defects. If we write $q=m{q_I} + 2{j_M}$
where ${q_I}=0,1$ and ${j_M}=0,1,\ldots,m-1$, then \cite{Lindner12}
\begin{equation}
U_{k,k+1} = e^{i\pi m {q_I^2}} \, \omega^{j_M^2}
\end{equation}
where $\omega=e^{2\pi i/m}$. The first factor is the braiding transformation
for Ising anyons if $m\cong 1\!\pmod{4}$ and for the opposite-chirality
version of Ising anyons if $m\cong 3\!\pmod{4}$. The second factor
can be rewritten using the Gauss quadratic sum as:
\begin{equation}
\omega^{j_M^2} = \frac{1}{\sqrt{m}}\sum_{j=0}^{m-1} \omega^{j^2} {u_k^j}
\end{equation}
which is the same, up to a phase, as Eq. (\ref{eqn:braid-group-rep})
(see also Eq. 25 of Ref. \onlinecite{Lindner12}).

Thus, these physical models give a very natural interpretation to the
elements of the extra special group $H(2n-1,\mathbb{F}_{m})$: these are the operators
that rotate the phase of the superconducting order parameter or the
ferromagnetic spin by $4\pi$. Their eigenvalues are just the allowed
charges/spins on gapped intervals modulo charge $2e$ or spin-$1$.

However, it is also important to note the differences between the metaplectic-Majorana
TQFT and the models of Refs. \onlinecite{Clarke12,Lindner12,Cheng12,Vaezi12}.
The latter models are gapless since they have the Goldstone boson associated with
superconductivity (which is not given a gap by the coupling to a 3D electromagnetic
field). Therefore, these models are, at best, in quasi-topological phases \cite{Bonderson12}
and are related to the metaplectic TQFT in the same way that
chiral $p$-wave superconductors are related to Ising anyons: they have some but not
all of the properties of a true topological phase. Furthermore, we note that
the models of Refs. \onlinecite{Clarke12,Lindner12,Cheng12,Vaezi12}
do not appear to have a $Z$ particle. They have $2n$-particle
Hilbert spaces of dimension $(2m)^{n-1}$. This is the same as the direct sum
${\cal H}^{I}_{2n}\oplus{\cal H}^{Z}_{2n}$, which suggests that these models
do not distinguish between the $Z$ particle and the vacuum.
Moreover, the $Y_i$ particles are non-Abelian in the metaplectic-Majorana TQFT,
but the charges/spins are Abelian anyons in the models of
Refs. \onlinecite{Clarke12,Lindner12,Cheng12,Vaezi12} .
In the metaplectic-Majorana TQFT,
when a $Y_j$ particle is taken around a $Y_k$ particle,
a phase $e^{\pm i\pi j k/m}$ results, depending on whether
they fuse to $Y_{|j-k|}$ or $Y_{\text{min}(i+j,m-i-j)}$. Each of these
possibilities occurs twice (for each pair) if we allow the total charge to be $I$ or $Z$.
In the models of Refs. \onlinecite{Clarke12,Lindner12,Cheng12,Vaezi12},
however, the phase $e^{i\pi j k/m}$ results when a charge
$j$ is taken around charge $k$ or $m-j$ is taken around $m-k$
while $e^{-i\pi j k/m}$ results when a charge
$j$ is taken around charge $m-k$ or $m-j$ is taken around $k$.
In our model, we can only determine the phase resulting from a braid
by performing a measurement of the total topological charge
of the two particles. In the models of Refs. \onlinecite{Clarke12,Lindner12,Cheng12,Vaezi12},
however, we can determine the phase resulting from a braid
by simultaneously measuring the charges of the two intervals.

A possible path to understanding the relation between the metaplectic-Majorana
TQFT and the models of Refs. \onlinecite{Clarke12,Lindner12,Cheng12,Vaezi12}
is through Slingerland and Bais' \cite{Slingerland09} analysis of $SU(2)_4$,
which is equivalent to $SO(3)_2$. They show that
the condensation of the spin-$2$ particle (the $Z$ particle),
causes the confinement of the spin-$1/2$ and $3/2$ particles
(the $X$ and $X'$ particles). The $Y_1$ particle splits into $2$
particles which, together with $I$, form a $\mathbb{Z}_3$ multiplet.
A version of this scenario should occur for general $SO(m)_2$, and
may be related to the models of Refs. \onlinecite{Clarke12,Lindner12,Cheng12,Vaezi12}:
the charges/spins on intervals are the Abelian quasiparticles of the theory,
which are the only `true' quasiparticles in the theory since they are not confined,
while $X$ particles are confined but, if the energy required to pull them apart
is supplied, then a projective remnant of their non-Abelian braiding properties
survives. The dislocations of Ref. \onlinecite{Barkeshli11b,Barkeshli12b} may have a
similar relation to the $X$ particles of the metaplectic TQFT.

\section{Discussion}

It was recently realized that the transformations associated with Ising
anyons could also be realized in three spatial dimensions \cite{Teo10,Freedman11a}.
Although there is no braiding in three dimensions, extended objects,
which could be viewed as particles connected to ribbons, would have the
topology of their configuration space governed by an enhancement of the permutation
group, $E(\mathbb{Z}_2^{2n-1}\rtimes S_{2n})$ (here, the $E(...)$ denotes the restriction
to elements whose combined parity is even). The $\mathbb{Z}_2$ factors keep track
of the twisting of the ribbons, modulo a $4\pi$ twist, which can be undone.
Solitons supporting Majorana zero modes realize a projective representation of this
group, which has image $H(n-2,{\mathbb{F}_2})\rtimes S_{2n}$. Thus, the non-Abelian
statistics of Ising anyons can be understood as simply permutations together
with $2\pi$ ribbon twists of pairs of particles. Two such twists anti-commute if they
share a particle (but not both). The non-Abelian statistics of $X$ particles
in $SO(m)_2$ is a generalization of this to fractional twists: $H(n-2,{\mathbb{F}_2})$
is replaced by $H(n-2,{\mathbb{F}_m})$ so that the (purely fictitious) ribbons
connecting particles can be twisted up to $m-1$ times.

Although the resulting unitary transformations are richer than those
of Ising anyons, this TQFT is still incapable of performing universal
quantum computation through braiding alone. The braid group
has an image which is finite. However, a certain link invariant associated
with the amplitude for creating pairs of $Y_1$ particles, braiding them,
and annihilating them in pairs is $\# P$-hard to compute. This suggests that
there may be greater computational power lurking just beneath the surface
of this theory and, perhaps, that it becomes apparent when braiding is
supplemented by measurement at intermediate steps of a computation.
Specific protocols by which universal quantum computation could be achieved
with metaplectic anyons (with or without Majorana zero modes) are an interesting
open problem.

\acknowledgements
We would like to thank J. Alicea, E. Berg, P. Bonderson, M. Cheng,
D. Clarke, B. Conrad, N. Lindner, K. Shtengel, and J. Yard for discussions.
M.B.H. is partially supported by a Simons Investigator
award from the Simons Foundation.
C.N. is supported by the DARPA QuEST program
and the AFOSR under grant FA9550-10-1-0524.
Z. W. is partially supported by NSF DMS 1108736.

\appendix

\section{$Sp(2n,{\mathbb{F}_m})$ and the braid group of metaplectic anyons}
\label{sec:metaplectic}

In this appendix, we will discuss in greater detail the image of the representation of the
braid group associated with $X$-particles.
We begin with a $2n$-dimensional vector space $V_{2n}$ over $\mathbb{F}_m$
(with $m$ assumed to be prime) equipped with a
non-degenerate symplectic form $[,]$. We can take as a basis of this
vector space ${v_i}=(0,\ldots,0,1,0\ldots,0)$, which has zero for every entry
except for the $i^{\rm th}$, which is $1$. We will take the symplectic form to
be $\left[{v_i},{v_j}\right]=\pm\delta_{i\pm 1,j}$. The group of linear transformations that
preserve the symplectic form $[,]$ is the symplectic group
$Sp(2n,{\mathbb{F}_m})$. This is a finite group whose order can be determined as follows.
We want all ways of choosing ${v_1},\ldots,v_{2n}$ so that
$\left[{v_i},{v_j}\right]=\pm\delta_{i\pm 1,j}$. There are $m^{2n}$ vectors in $V_{2n}$
since it is composed of all linear combinations of ${v_1},\ldots,v_{2n}$ with
coefficients in $\mathbb{F}_m$. Therefore, there are $m^{2n}-1$ ways to
choose ${v_1}\neq 0$. There is a $(2n-1)$-dimensional space of vectors
$v$ with $\left[{v_1},{v}\right]=0$. Therefore, there are $m^{2n}-m^{2n-1}$
choices of vector $v_2$ with $\left[{v_1},{v_2}\right]\neq 0$. Since
the possible non-zero values of $\left[{v_1},{v_2}\right]$ are
$1,2,\ldots,m-1$, there are $(m^{2n}-m^{2n-1})/(m-1) = m^{2n-1}$ choices
of $v_2$ with $\left[{v_1},{v_2}\right]=1$. Continuing in this way, we
find that there are
\begin{equation}
\prod_{i=1}^{n} \left(m^{2i}-1\right)m^{2i-1} =
m^{n^2} \prod_{i=1}^{n} \left(m^{2i}-1\right)
\end{equation}
elements of the group $Sp(2n,{\mathbb{F}_m})$.

Now consider $V_{2n}$ as an additive group.
Consider a central extension $G$:
$1\rightarrow \mathbb{F}_m \rightarrow G \rightarrow V_{2n} \rightarrow 1$.
Since $V_{2n}$ is Abelian, the commutator map $G \times G\rightarrow G$
given by $({g_1},{g_2})\rightarrow {g_1}{g_2}{g_1}^{-1}{g_2}^{-1}$
takes values in the center $\mathbb{F}_m$ and
is unaffected by multiplication by the center,
so it defines a map $V_{2n}\times V_{2n}\rightarrow\mathbb{F}_m$.
In the case of the specific central extension that is usually called
the `extra special group' or `Heisenberg group', which we denote by $H(2n,{\mathbb{F}_m})$,
this map is just the symplectic form $[,]$. The elements of $H(2n,{\mathbb{F}_m})$
can be written in the form $(v,k)$, where $v\in V_{2n}$ and $k\in\mathbb{F}_m$.
The multiplication rule is $({v_1},{k_1})\cdot({v_1},{k_1})=({v_1}+{v_2},{k_1}+{k_2}+[{v_1},{v_2}])$.
For the basis taken above with $\left[{v_i},{v_j}\right]=\pm\delta_{i\pm 1,j}$,
if write ${u_i}\equiv ({v_i},0)$ and $z\equiv (0,1)$, then we have the defining relations
introduced in Sec. \ref{sec:braid-group-rep}
\begin{eqnarray}
\label{eqn:extra-special-again}
{u_i^m} &=& 1\, ,\,\,{z^m} \,= \, 1 \cr
u_{i} u_{i+1} &=& z u_{i+1} u_{i}\cr
u_{i} u_{j} &=& u_{j} u_{i} \, , {\hskip 0.25 cm} |i-j|>1\cr
u_{i} z &=& z u_{i} \, .
\end{eqnarray}
If we, instead, take a basis $f_i$ of $V_{2n}$ with
$\left[f_{2i-1},f_{2j}\right]=\delta_{ij}$ and
$\left[f_{2i-1},f_{2j-1}\right]=\left[f_{2i},f_{2j}\right]=0$,
then we have a different generating set for $H(2n,{\mathbb{F}_m})$:
${X_i}\equiv (f_{2i-1},0)$, ${Z_i}~\equiv~(f_{2i},0)$, $z\equiv (0,1)$
satisfying:
\begin{eqnarray}
\label{eqn:Heisenberg-Z-X}
X_{i} X_{j} &=& X_{j} X_{i} \, ,\,\,  Z_{i} Z_{j} = Z_{j} Z_{i}\cr
X_{i} Z_{j} &=& \, z^{\delta_{ij}}\,Z_{j} X_{i}\cr
X_{i} z &=& z X_{i} \, ,\,\,  Z_{i} z = z Z_{i}
\end{eqnarray}
These two presentations of $H(2n,{\mathbb{F}_m})$ are related
by $u_{2i-1}=X_{i}$, $u_{2i}=Z_{i}Z^\dagger_{i+1}$ for $i\neq n$
and $u_{2n}=Z_{n}$.

The symplectic group $Sp(2n,{\mathbb{F}_m})$ of $V_{2n}$ acts on
$H(2n,{\mathbb{F}_m})$ in the natural way. These are automorphisms that act trivially
on the center $Z(H(2n,{\mathbb{F}_m}))$ of $H(2n,{\mathbb{F}_m})$. In addition, the inner automorphisms -- conjugation by elements of $H(2n,{\mathbb{F}_m})$ --
are also trivial on $Z(H(2n,{\mathbb{F}_m}))$.
In fact, the group of automorphisms of $H(2n,{\mathbb{F}_m})$
that are trivial on $Z(H(2n,{\mathbb{F}_m}))$
is given by $Sp(2n,{\mathbb{F}_m})\ltimes V_{2n}$. ($V_{2n}$, rather than
$H(2n,{\mathbb{F}_m})$, is the second factor in this semi-direct product because
$Z(H(2n,{\mathbb{F}_m}))$ acts trivially on $H(2n,{\mathbb{F}_m})$ by conjugation, so only
$H(2n,{\mathbb{F}_m})/Z(H(2n,{\mathbb{F}_m}))=V_{2n}$ appears).
The group $Sp(2n,{\mathbb{F}_m})\ltimes H(2n,{\mathbb{F}_m})$ is, therefore, an extension
of the group of automorphisms of $H(2n,{\mathbb{F}_m})$
that are trivial on $Z(H(2n,{\mathbb{F}_m}))$;
the group has been extended by $Z(H(2n,{\mathbb{F}_m}))$.

This is a useful extension to consider because, given an irreducible
representation $M$ of $H(2n,{\mathbb{F}_m})$, there is a unique induced representation
$X$ of $Sp(2n,{\mathbb{F}_m})\ltimes H(2n,{\mathbb{F}_m})$ whose restriction to
$H(2n,{\mathbb{F}_m})$ is $M$, as shown in Ref. \onlinecite{Goldschmidt89}
and as we discuss in the next paragraph.
Moreover, given a representation $\lambda(k)=\omega^k$ of
$Z(H(2n,{\mathbb{F}_m}))$, there is a unique induced representation $M$ of
$H(2n,{\mathbb{F}_m})$ whose restriction to its center is $\lambda(k)$.
Here, $\omega$ is an $m^{\rm th}$ root of unity.
Let $M_v\equiv M(v,0)$ for $(v,0)\in H(2n,{\mathbb{F}_m})$.
Then, the induced representation of $H(2n,{\mathbb{F}_m})$
must satisfy ${M_u}{M_v}=\lambda([u,v])M_{u+v}$.
Consequently, $M_{v_i}^m=1$, $M_{v_i}M_{v_{i+1}}=\omega^{-2}M_{v_{i+1}}M_{v_i}$,
$M_{v_i}M_{v_{j}}=M_{v_{j}}M_{v_i}$ for $|i-j|>1$.

This representation of $H(2n,{\mathbb{F}_m})$ induces a representation of
$Sp(2n,{\mathbb{F}_m})\ltimes H(2n,{\mathbb{F}_m})$ as follows.
Consider the action of $g\in Sp(2n,{\mathbb{F}_m})$ on $h\in H(2n,{\mathbb{F}_m})$
by conjugation inside $Sp(2n,{\mathbb{F}_m})\ltimes H(2n,{\mathbb{F}_m})$:
$h\rightarrow ghg^{-1}$. Since $H(2n,{\mathbb{F}_m})$ is a normal
subgroup of $Sp(2n,{\mathbb{F}_m})\ltimes H(2n,{\mathbb{F}_m})$,
$ghg^{-1} \in H(2n,{\mathbb{F}_m})$. Therefore, for each $g\in Sp(2n,{\mathbb{F}_m})$
there is a representation of $H(2n,{\mathbb{F}_m})$ given by
$h\rightarrow M_{ghg^{-1}}$. But since there is a unique representation,
there must be a unitary transformation $X(g)$ such that
$M_{ghg^{-1}} = X(g) {M_h} X(g)^{-1}$. This defines $X(g)$
up to a scalar. In fact, $X(g)$ is not quite a linear representation
of $Sp(2n,{\mathbb{F}_m})$. It is a projective representation or, equivalently,
it is a linear representation of the double-cover of $Sp(2n,{\mathbb{F}_m})$,
namely the {\it metaplectic group}.
This representation can be given explicitly 
in terms of the $M_v$ according to the relation:
\begin{equation}
\label{eqn:induced-rep}
X(g) = \sum_{v\in {V_1}(g)} {a_v}(g) M_v
\end{equation}
where ${V_1}(g)=\text{im}(1-g)$. It may further be shown that
${a_v}(g)=\lambda([u,g(u)]) \,{a_0}(g)$ where
$v=u-g(u)\in {V_1}(g)$.

We now consider the following map \cite{Goldschmidt89} from
$B_{2n+1}\rightarrow Sp(2n,{\mathbb{F}_m})$. To the generator $\sigma_i$
of $B_{2n+1}$, we associate the $Sp(2n,{\mathbb{F}_m})$ transformation
$\hat{\sigma}_i$ that acts on $V_{2n}$ according to
\begin{eqnarray}
\label{eqn:braid-symplectic}
{\hat{\sigma}_i}({v_i}) &=& {v_i}\cr
{\hat{\sigma}_i}(v_{i\pm 1}) &=& v_{i\pm 1} \mp {v_i}\cr
{\hat{\sigma}_i}(v_{j}) &=& {v_j} \, , {\hskip 0.25 cm} |i-j|>1
\end{eqnarray}
It may be directly checked that this transformation preserves
the symplectic form $[,]$ and that $\hat{\sigma}_i$ satisfy the
defining relations of the braid group. Then, from Eq. \ref{eqn:induced-rep},
there is a braid group representation
\begin{equation}
\label{eqn:bg-induced-rep}
X({\sigma_i}) = \sum_{v\in {V_1}({\sigma_i})} {a_v}({\sigma_i}) M_v
\end{equation}
From Eq. (\ref{eqn:braid-symplectic}), we see that
${V_1}({\sigma_i})=\{k{v_i}| k\in \mathbb{F}_m\}$. Hence,
for $g={\sigma_i}$, we have $v=k{v_i}\in {V_1}({\sigma_i})$
and $u=kv_{i+1}$ such that $v=u-g(u)$. Consequently,
${a_v}({\hat{\sigma}_i})=\lambda([u,{\hat{\sigma}_i}(u)]) \,{a_0}({\hat{\sigma}_i})
=\lambda([kv_{i+1},-kv_{i}]){a_0}({\sigma_i})=\omega^{k^2}{a_0}({\sigma_i})$
Therefore,
\begin{equation}
X({\sigma_i}) = {\cal N} \sum_{k=0}^{m-1} \omega^{k^2} M_{kv_i}
= {\cal N} \sum_{k=0}^{m-1} \omega^{k^2} M_{v_i}^k
\end{equation}
where ${\cal N}$ is a normalization constant.
We see that this is the same as the braid group
representation in Eq. (\ref{eqn:braid-group-rep})
which determines the braiding of $2n+1$ $X$-particles.
Therefore, the image of the braid group representation
of $2n+1$ $X$-particles is equal to the metaplectic
representation of $Sp(2n,{\mathbb{F}_m})$.

The technical reason why the case of $2n+1$ particles
is simple is that the braid group $B_{2n+1}$ has an even number
of generators ${\sigma_1}, \ldots, \sigma_{2n}$ (since $\sigma_i$
exchanges particles $i$ and $i+1$).
For an even number of generators, there is a natural mapping to
$Sp(2n,{\mathbb{F}_m})$ since the latter is defined on a
symplectic vector space, which must be even-dimensional.
For an even number $2n$ of particles, the braid group $B_{2n}$
has an odd number of generators. In order to construct the
corresponding symplectic group, we begin with the
symplectic vector space $V_{2n}$ over ${\mathbb{F}_m}$
and pick a vector $e_1 \in V_{2n}$.
Then we consider the group $G$ of linear transformations that
preserve the symplectic structure $[,]$ on $V_{2n}$ {\it and}
leave $e_1$ invariant. The vector space orthogonal to
$e_1$ is $(2n-1)$-dimensional, so $G$ is the odd-dimensional analogue
of a symplectic group and is sometimes called an {\it odd symplectic group} \cite{Gelfand84}.
Clearly, $Sp(2n-2,{\mathbb{F}_m})\subset G$. The rest of $G$
is given by transformations of the following form. Let $e_{2n}$ be the vector that
satisfies $\left[{e_1},e_{2n}\right]=1$. Then, for any $v\in \text{span}({e_2},\ldots,e_{2n-1})$
and $k\in \mathbb{F}_m$, the symplectic form $[,]$ and $e_1$
are left invariant by the transformations $e_{2n}\rightarrow e_{2n} + v + k{e_1}$
and ${e_i}\rightarrow {e_i} + [v,{e_i}] {e_1}$ for $i=2,3,\ldots,2n-1$.
These transformations, parametrized by $(v,k)$ form the group
$H(2n-2,m)$, as discussed above. They can be written explicitly
in matrix form as
\begin{equation}
\begin{pmatrix}
e_{2n}\\
e_{2n-1}\\
\vdots\\
{e_2}\\
{e_1}
\end{pmatrix} \rightarrow
\begin{pmatrix}1 & a^T_{n-1} & -b^T_{n-1} & c\\
0 & I_{n-1} & 0 & b_{n-1}\\
0 & 0 & I_{n-1} & a_{n-1}\\
0 & 0 & 0  & 1
 \end{pmatrix}
\begin{pmatrix}
e_{2n}\\
e_{2n-1}\\
\vdots\\
{e_2}\\
{e_1}
\end{pmatrix}
\end{equation}
where $a_{n-1}, b_{n-1}$ are $(n-1)$-component column vectors
over $\mathbb{F}_m$, $c\in \mathbb{F}_m$, $ I_{n-1}$ is
the $(n-1)\times(n-1)$ identity matrix, and the basis ${e_2}, {e_3}, \ldots e_{2n-1}$
is chosen so that $[{e_i},e_{2n+1-i}]=1$ for $i\leq n$ and $[{e_i},e_{j}]=0$ for
$j\neq 2n+1-i$. This is precisely the group $H(2n-2,m)$ in its representation
as upper triangular matrices. Then, following the steps given above for an odd number of particles,
we obtain a mapping $B_{2n}\rightarrow Sp(2n-2,{\mathbb{F}_m})\ltimes H(2n-2,m)$.

\bibliographystyle{prsty}
\bibliography{twisted-Majorana}

\end{document}